\documentclass[aps,prl,twocolumn,groupedaddress,floatfix]{revtex4}

\usepackage{graphicx}
\usepackage{dcolumn}
\usepackage{bm}

\newcommand{\Erms}{E_{\rm{rms}}}

\begin{document}


\title{Computational Study of Turbulent-Laminar Patterns in Couette Flow}

\author{Dwight Barkley}
\email[]{barkley@maths.warwick.ac.uk}
\homepage[]{www.maths.warwick.ac.uk/~barkley}
\affiliation{Mathematics Institute, University of Warwick, 
Coventry CV4 7AL, United Kingdom}

\author{Laurette S.\ Tuckerman}
\email[]{laurette@limsi.fr}
\homepage[]{www.limsi.fr/Individu/laurette}
\affiliation{LIMSI-CNRS, BP 133, 91403 Orsay, France}

\date{\today}

\begin{abstract}

Turbulent-laminar patterns near transition are simulated in plane Couette flow
using an extension of the minimal flow unit methodology.  Computational
domains are of minimal size in two directions but large in the
third. The long direction can be tilted at any prescribed angle to the
streamwise direction.  Three types of patterned states are found and studied:
periodic, localized, and intermittent.  These correspond closely to
observations in large aspect ratio experiments.

\end{abstract}

\pacs{47.20.-k, 47.27.-i, 47.54.+r, 47.60.+i}

\maketitle

\begin{figure}
\includegraphics[width=2.5in]{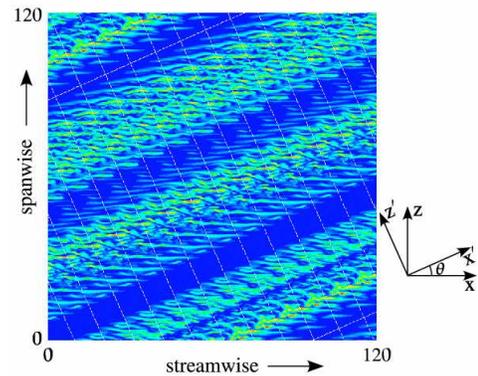}
\caption{Turbulent-laminar pattern at Reynolds number 350.  The computational
domain (outlined in white, aligned along $x^\prime$, $z^\prime$) is repeated
periodically to tile an extended region.  The kinetic energy is visualized in
a plane midway between and parallel to the plates moving in the streamwise
($x$) direction. Uniform gray or blue corresponds to laminar flow.  The sides
of the image are 60 times the plate separation $L_y=2$; the pattern wavelength
is 20 $L_y$.  Streamwise streaks (on the scale of $L_y$) are visible at the
edges of the turbulent regions.
\label{fig:visualization}}
\end{figure}

Plane Couette flow -- the flow between two infinite parallel plates moving in
opposite directions -- undergoes a subcritical (discontinuous) transition from
laminar flow to turbulence as the Reynolds number is increased.  Due to its
simplicity, this flow has long served as one of the canonical examples for
understanding shear turbulence and the subcritical transition process typical
of channel and pipe flows
\cite{Jimenez,Lundbladh,Daviaud,Tillmark,Hamilton,Dauchot,Schmiegel,Schumacher,Waleffe,Hof,Faisst,Manneville}.
Only recently was it discovered in very large aspect ratio experiments by
Prigent {\em et al.}~\cite{Prigent,Prigent2,Prigent3} that this flow also
exhibits remarkable pattern formation near transition.
Figure~\ref{fig:visualization} shows such a pattern, not from experiment, but
from numerical computations reported here.  An essentially steady, spatially
periodic pattern of distinct regions of turbulent and laminar flow emerges
spontaneously from uniform turbulence as the Reynolds number is decreased.  It
now appears that turbulent-laminar patterns are inevitable intermediate states
on the route from turbulent to laminar flow in large aspect ratio plane
Couette flow.

Related patterns have a long history in fluid dynamics.  In Taylor-Couette
flow between counter-rotating cylinders, Coles~\cite{Coles} first discovered a
state known as spiral turbulence with coexisting turbulent and laminar
regions. This state was famously commented on by Feynman \cite{Feynman} and
has attracted attention as an example of a coherent structure comprising both
turbulence and long-range order \cite{vanAtta,Andereck,Hegseth,Mutabazi}.
Until recently all experimental studies of this state showed only one
turbulent and one laminar patch.  Prigent {\em et
al.}~\cite{Prigent,Prigent2,Prigent3} found that in a very large-aspect-ratio
Taylor-Couette system, the turbulent and laminar regions form a periodic
pattern, of which the original observations of Coles comprised only one
wavelength.  Cros and Le Gal \cite{Cros} discovered large-scale turbulent
spirals as well, in experiments on the shear flow between a stationary and a
rotating disk.  The Reynolds-number thresholds, wavelengths, and angles are
very similar for all of these turbulent patterned flows.  Moreover, Prigent
{\em et al.} suggest that the turbulent spots
\cite{Lundbladh,Daviaud,Tillmark,Dauchot,Bottin,Schumacher,Hof,Manneville,Cros}
long known to exist near transition are essentially a manifestation of the
same mechanism.

We report the first direct numerical simulation of turbulent-laminar patterns.
Our simulations are designed to reduce computational expense, to establish
minimal physical conditions necessary to produce these large-scale patterns,
and to impose and thereby investigate the pattern wavelength and orientation.

\begin{figure}
\includegraphics[width=3.0in]{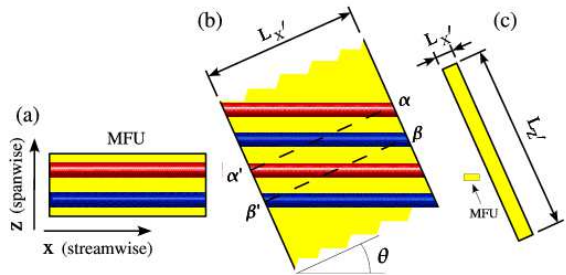}
\caption{Simulation domains.  The wall-normal direction $y$ is
not seen; $L_y=2$.  The gray or colored bars represent streamwise vortex pairs
with a spanwise spacing of 4.  (The vortices are schematic; these are dynamic
features of the actual flow.)  (a) MFU domain of size $10 \times 4$.  (b)
Central portion of a domain [on the same scale as (a)] titled to the
streamwise direction.  $\alpha$, $\alpha^\prime$ and $\beta$, $\beta^\prime$
are pairs of points identified under periodic boundary conditions in
$x^\prime$.  (c) Full tilted domain with  $L_{x^\prime} =10$, 
$L_{z^\prime} = 120$, $\theta=24^\circ$.  
On this scale the MFU domain, shown for comparison, is small.
\label{fig:domains}}
\end{figure}

Our study extends minimal-flow-unit (MFU) simulations of turbulent channel
flows~\cite{Jimenez,Hamilton,Waleffe} and we begin by recalling these.  The
plates located at $y=\pm 1$ move at unit velocities $\pm{\bf\hat{x}}$.  The
Reynolds number is $Re=1/\nu$, where $\nu$ is the kinematic viscosity of the
fluid.  The simple Couette solution ${\bf u_C}\equiv y{\bf\hat{x}}$ is linearly
stable for all values of $Re$. However, above a critical $Re$ near $325$
\cite{Dauchot}, transition to turbulence occurs for sufficiently large
perturbations. The turbulence is characterized by the cyclical generation and
breakdown of streaks by streamwise-oriented vortices with a natural spanwise
pair spacing of about 4~ \cite{Jimenez,Hamilton,Waleffe,Bottin,Barkley}.  In
the MFU approach, a periodic domain of minimal lateral dimensions is sought
which can sustain this basic cycle.  For plane Couette flow near transition,
the currently accepted~\cite{Waleffe} size is approximately $L_x \times L_z=
10 \times 4$ [Fig.~\ref{fig:domains}(a)].

We extend the MFU computations in two ways.  First we tilt the domain at angle
$\theta$ to the streamwise direction [Fig.~\ref{fig:domains}(b)].  We
designate by $x^\prime$ and $z^\prime$ the periodic directions of the tilted
domain.  To respect the spanwise streak spacing while imposing
periodic boundary conditions in $x^\prime$, the domain satisfies $L_{x^\prime}
\sin \theta \simeq 4$, for $\theta>0$.  (For $\theta=0$, we only require
$L_{x^\prime} \gtrsim 10$.)  Secondly, we greatly extend one of the dimensions
past the MFU requirement [Fig.~\ref{fig:domains}(c)].  In practice we use
$L_{z^\prime}$ between 30 and 220, usually 120.  We can thus simulate large
length scales oblique to the streamwise direction.

The incompressible Navier-Stokes equations are simulated using a
spectral-element ($x^\prime$-$y$) -- Fourier ($z^\prime$)
code~\cite{Henderson}.  The boundary conditions are no-slip at the moving
plates and periodic in the $x^\prime$ and $z^\prime$ directions.  The spatial
resolution for the $L_{x^\prime}\times L_y \times L_{z^\prime} = 10\times
2\times 120$ domain is $N_x \times N_y \times N_z = 61 \times 31 \times 1024$,
consistent with that previously used at these values of $Re$
\cite{Hamilton,Waleffe,Schumacher}.  Results have also been verfied at higher
resolutions.

We make two comments distinguishing our approach.  Experimentalists
\cite{Prigent,Prigent2,Prigent3} varied $Re$ and reported the properties of the resulting
patterns: in particular they measured angles and wavelengths varying from
$\theta=25^\circ$ and $\lambda_{z^\prime}=46$ at $Re=394$ to $\theta=37^\circ$
and $\lambda_{z^\prime}=60$ at $Re=340$.  (They extrapolated the domain of
existence to be $325 \leq Re \leq 415$.)  In contrast, we fix the pattern
angle and wavelength: in this way, we can determine the boundaries in
parameter space within which each pattern can exist.  Second, all the
turbulent states we report are bistable with simple Couette flow. A major
goal \cite{Dauchot,Schmiegel,Hof,Faisst}, not addressed here, has been the
determination of lifetimes and transition probabilities of turbulent flow as a
function of amplitude and $Re$.

\begin{figure}
\includegraphics[width=2.8in]{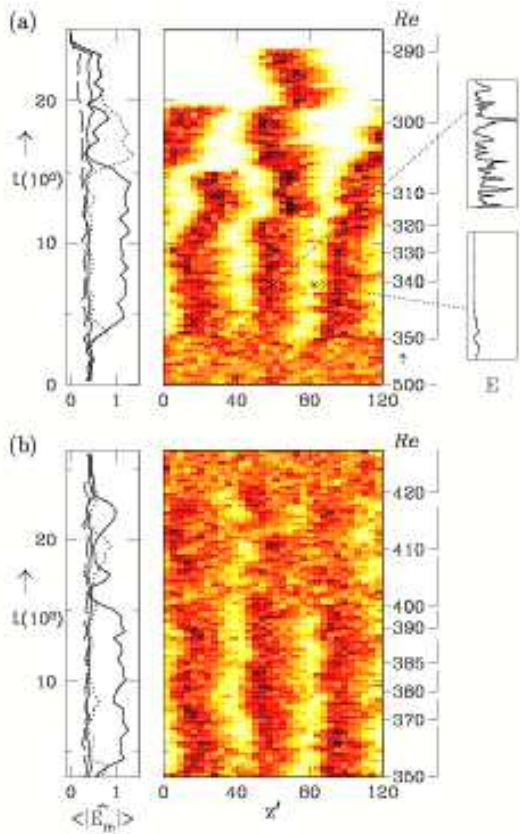} 
\caption{Space-time evolution of turbulent-laminar patterns in the domain
$L_{x^\prime} \times L_{z^\prime} = 10 \times 120$, $\theta = 24^\circ$.  Time
evolves upward with changes in $Re$ indicated on the right.  Grayscale or
color plots: kinetic energy fluctuations $\Erms$ on a space-time grid.  The
same scale is used for all space-time plots, with $\Erms = 0$ in white.
Insets: kinetic energy plotted over a time window $T=250$ in a turbulent and
laminar region.  Left: Spectral peaks in the averaged spatial Fourier
transform of kinetic energy with $m=3$ (solid) and $m=2$ (dotted), 0
(long-dashed), and 1 (short-dashed).  (a) States seen upon decreasing $Re$, 
from uniform turbulence at $Re=500$, through various patterned states, 
ending in simple Couette flow at $Re=290$.  
(b) States seen upon increasing $Re$, from the
three-banded laminar-turbulent pattern at $Re=350$ to uniform turbulence at
$Re=420$. }
\label{fig:spacetime}
\end{figure}

We begin with simulations exploring the dependence of patterns on $Re$.  To
allow the system sufficient freedom to select different states, we set
$L_{z^\prime} = 120$, two to three times the experimentally observed
wavelength. We fix $\theta=24^\circ$, near its observed value at pattern
onset.  Figures~\ref{fig:spacetime}(a) and (b) show two long series of
simulations spanning the range $290 \leq Re \leq 500$.  Space-time diagrams
are shown for decreasing and increasing $Re$ in discrete steps over time.  In
each case, kinetic energy fluctuations are on the right and principle
peaks in associated spatial Fourier transforms are on the left.

More specifically, we compute $E = |{\bf u-u_C}|^2/2$ at 32 points equally
spaced in $z^\prime$ along a line ($x^\prime=y=0$) in the mid-channel.  We
compute $\Erms$, the rms of $E$ in time windows of size $T=250$.  This gives a
measure of the flow's turbulent intensity on a space-time grid.  (Other
measures such as the rms of individual velocity components gives similar
results.)  Time windows in Fig.~\ref{fig:spacetime}(a) show $E$ from which
$\Erms$ is computed at two points on the space-time grid.  For the spectra on
the left, we compute the instantaneous spatial Fourier transform $\hat{E}_m$
of $E$ for the same 32 points in the mid-channel. We take the modulus (to
remove phase information) and average over windows of 
length $T=500$ to obtain $\langle|\hat{E}_m|\rangle$.

In Fig.~\ref{fig:spacetime}(a) a turbulent flow is initialized at $Re=500$ by
perturbing simple Couette flow.  We call the resulting unpatterned state {\it
uniform turbulence}.  Its spectrum is flat.  $Re$ is decreased quickly to
$350$ where a pattern forms with three distinct turbulent and laminar
regions. The $m=3$ spectral peak emerges.  The selected wavelength of 40 agrees
closely with experiment \cite{Prigent,Prigent2,Prigent3}.  
$Re$ is kept at 350 long enough to
show that this pattern is stable.  The final flow at $Re=350$ is visualized in
Fig.~\ref{fig:visualization}.  The pattern remains qualitatively the same
through $Re=320$.  $\Erms$ is systematically greater to the left of the band
center.  (Note that, due to the imposed tilt, there is no reflection symmetry
in $z^\prime$.)  At $Re=310$ the pattern loses one turbulent region,
accompanied by the emergence of the $m=2$ spectral peak. At $Re=300$, a single
turbulent region remains, and finally, at $Re=290$, the flow reverts to simple
Couette flow.

Figure~\ref{fig:spacetime}(b) shows states obtained by increasing $Re$
starting from $Re=350$. The steady three-banded pattern persists up through
$Re=390$.  At $Re=400$ and $410$ the pattern is no longer steady: bands are
less well defined and laminar regions appear and disappear (see
below). Uniform turbulence is obtained at $Re=420$.

\begin{figure}
\includegraphics[width=2.8in]{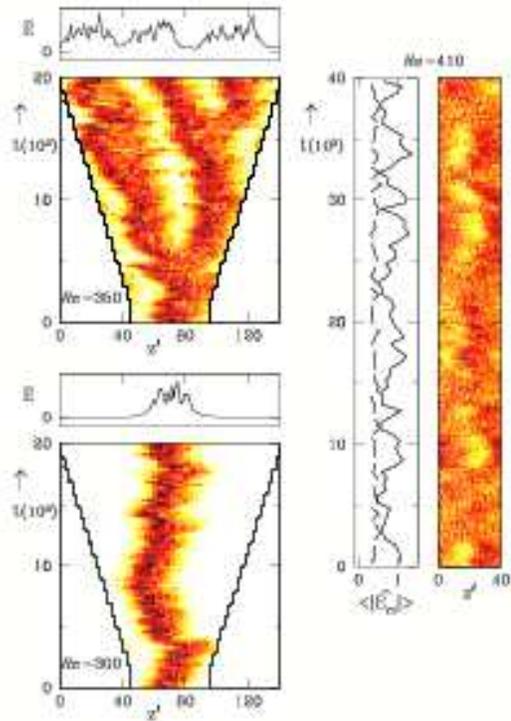}
\caption{Simulations at $Re=350$, $Re=300$, and $Re=410$ illustrating three
distinct states: periodic, localized, intermittent.  
Space-time representation of $E_{\rm{rms}}$ is as in Fig.~\ref{fig:spacetime}.
For $Re=350$ and $Re=300$ the domain length is increased from
$L_{z^\prime}=50$ to $L_{z^\prime}=140$ in increments of 5.  The integrated
energy profile $\bar E(z^\prime)$ is shown at the final time.  For $Re=410$ a
single long simulation is shown for $L_{z^\prime}=40$, accompanied by $m=1$
(solid) and $m=0$ (dashed) spectral peaks. }
\label{fig:growit}
\end{figure}

We now present evidence that the patterns in Fig.~\ref{fig:spacetime}
represent three qualitatively different states.  The banded state at $Re=350$
is fundamentally {\em spatially periodic}. To support this we show in
Fig.~\ref{fig:growit} a simulation at $Re=350$ in a domain whose length
$L_{z^\prime}$ is slowly increased.  The pattern adjusts to keep the
wavelength in the approximate range $35-65$ by splitting the turbulent bands
when they grow too large.  The instantaneous integrated energy profile $\bar E
\equiv \int\,dx^\prime\, dy \: E(x^\prime,y,z^\prime,t)$ is plotted at the
final time. Between the turbulent bands, $E$ does not reach zero and the flow,
while basically laminar, differs from the simple Couette solution $y{\bf
e_x}$.

\begin{figure}
\includegraphics[width=2.8in]{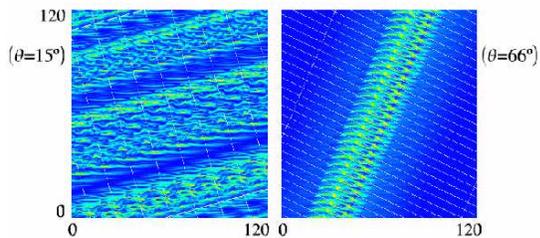}
\caption{Turbulent-laminar patterns at minimum ($\theta=15^\circ$) and maximum
($\theta=66^\circ$) angle for which they have been computed at $Re=350$.
Conventions as in Fig.~\ref{fig:visualization}.
\label{fig:2square}}
\end{figure}

In sharp contrast, the single turbulent patch seen in
Fig.~\ref{fig:spacetime}(a) prior to return to laminar Couette flow is a {\it
localized state}.  Figure~\ref{fig:growit} shows that in a domain of
increasing size at $Re=300$ a single turbulent region of approximately fixed
extent persists, independent of $L_{z^\prime}$.  Moreover, $\bar E$ decays to
zero exponentially as the flow approaches the simple Couette solution away
from the patch.  The localized states in our computations necessarily take the
form of bands when visualized in the $x-z$ plane [e.g.,
Fig.~\ref{fig:2square}(d) below].  Isolated bands and spots are reported
experimentally \cite{Prigent,Prigent2,Prigent3} near these values of $Re$.

The third behavior is displayed by the {\it intermittent state} in
Fig.~\ref{fig:spacetime} near the transition to uniform turbulence.
Figure~\ref{fig:growit} shows a long simulation at $Re=410$ in a domain
$L_{z\prime} = 40$.  The flow never stabilizes but instead quasi-laminar
regions nucleate and disappear continually.  The range of $\Erms$ in the
space-time plot is noticeably smaller than for the stable patterns.
Simulations with $L_{z^\prime} = 60$ show similar behavior.  These states have
been interpreted in~\cite{Prigent,Prigent2,Prigent3} as resulting from
noise-driven competition between banded patterns at equal and opposite angles.
However, the intermittency is captured in our simulations, even though the
competition between states of opposite angles is absent.

We have increased and decreased $Re$ gradually at $L_{z^\prime} = 40$ and
$L_{z^\prime} = 60$ and find no hysteresis in any of the transitions between
the turbulent states.

We have explored regions of existence for various states as a function of
$Re$, wavelength, and tilt.  By varying $L_{z^\prime}$ at $\theta=24^\circ$,
$Re=350$, we have determined that the minimum wavelength is 35 and the maximum
is 65.  For $L_{z^\prime} \lesssim 30$, uniform turbulence is obtained.  For
$L_{z^\prime} \gtrsim 70$ two bands of wavelength $L_{z^\prime}/2$ form (as in
Fig.~\ref{fig:growit}).  This range of allowed wavelengths is nearly
independent of $Re$ wherever we have been able to compute banded states.
Figure~\ref{fig:2square} shows a banded state at $\theta=15^\circ$ and a
localized state at $\theta=66^\circ$, the minimum and maximum angles for which
we have thus far obtained patterns for $L_{z^\prime}=120$, $Re=350$.  These
extreme states may not be stable without the imposed periodicity of the
computations.  The sequence of states seen for increasing $\theta$ at $Re=350$
is qualitatively the same as that for decreasing $Re$ at $\theta=24^\circ$.
At $\theta=0^\circ$ and $\theta=90^\circ$ we do not find patterns, but only
either uniform turbulence or simple Couette flow, with transition boundaries
$Re \approx 300$ for $\theta=0^\circ$ and $Re\approx 390$ for
$\theta=90^\circ$.  Full details will be reported elsewhere.

In past years minimal-flow-unit simulations have been used to great effect in
understanding shear turbulence. We have shown that the same philosophy can be
employed in the study of large-scale structures formed in turbulent flows.
Specifically, we have reported the first simulations of turbulent-laminar
patterns in plane Couette flow by numerically solving the Navier-Stokes
equations in domains with a single long direction.  The other dimensions are
just large enough to resolve the inter-plate distance and to contain an
integer number of longitudinal vortex pairs or streaks.  Thus we have
demonstrated that the patterns are quasi one-dimensional and we have
identified what we believe to be near-minimal conditions necessary for their
formation.  Key is that the computational domain be tilted obliquely to the
streamwise direction of the flow, otherwise no patterns are observed.  We have
found periodic, localized, and intermittent states where similar states are
observed experimentally.  We have explored the patterns' dependence on
Reynolds number and on imposed wavelength and tilt.  The existence of
localized states in our simulations is particularly interesting because this
suggests that the basic physics of isolated turbulent spots can be captured
without simulating two large lateral directions.  Future studies of
these states may shed light on the mechanisms responsible for
laminar-turbulent patterns and for turbulent transition.


We thank Olivier Dauchot for valuable discussions and Ron Henderson for
the use of {\tt Prism}.  We thank the CNRS and the
Royal Society for supporting this work.  The two CPU decades of computer time
used for this research were provided by the IDRIS-CNRS supercomputing center
under project 1119, and by the University of Warwick Centre for Scientific
Computing (with support from JREI grant JR00WASTEQ).

\bibliography{prl8}

\end{document}